# Mössbauer experiments in a rotating system: The so-called "synchronization effect" to explain the extra energy shift between emitted and absorbed radiation constitutes a complete failure

Alexander Kholmetskii[1], Tolga Yarman[2], Ozan Yarman[3] and Metin Arik [4]

[1]Department of Physics, Belarus State University, Minsk, Belarus; alkholmetskii@gmail.com
[2]Okan University, Istanbul, Turkey & Savronik, Eskisehir, Turkey; tyarman@gmail.com
[3]Istanbul University, Istanbul, Turkey; ozanyarman@ozanyarman.com
[4]Bogazici University, Istanbul, Turkey; metin.arik@boun.edu.tr

**Abstract:** We show that a new attempt by C. Corda to once more rehash his so-called "synchronization effect" in order to account for the origin of the extra energy shift between emitted and absorbed radiation in Mössbauer rotor experiments (C. Corda, Int. J. Mod. Phys. D, doi: 10.1142/S0218271819501311) is yet again erroneous, just as were his previous attempts (Ann. Phys. **355**, 360 (2015); Ann. Phys. **368**, 258 (2016); Int. J. Mod. Phys. D **27**, 1847016 (2018)). The correct approach presented herein with regards to the calculation of the energy shift between emitted and absorbed radiation in a rotating system leads to, as a matter of fact, no specific "synchronization effect".

## 1. Introduction

Recent Mössbauer experiments in a rotating system carried out by our team (see, e.g. [1-4]) have definitely confirmed the presence of an extra energy shift (next to the usual relativistic dilation of time) between emitted and a received radiation, which is expressed by the fact that, in the equation written for the transverse relative energy shift between emitted and received radiation in a rotating system

$$\frac{\Delta E}{E}=-k\frac{u^2}{c^2}, \tag{1}$$

the coefficient $k$ substantially exceeds the relativistic prediction 0.5 corresponding to the usual time dilation effect for the moving receiver.

We remind that in our experiments [1-4], which were originally stimulated by the prediction of T. Yarman [5], as well as in our subsequent re-analysis of the experiment by W. Kündig [6] revealing the inequality $k>0.6$ [7], we have found

$$k=0.66\pm0.03 \text{ [1, 2]} \tag{2}$$

and

$$k=0.69\pm0.02 \text{ [3, 4]}. \tag{3}$$

It must be emphasized that, in the measurement results (2) and (3), the deviation from the relativistic prediction $k=0.5$ exceeds by several times the measurement uncertainty, so much so that the presence of the extra energy shift (hereinafter abbreviated as EES) has definitely been confirmed. As a consequence, its physical explanation has since then became a strong necessity.

One of the follow-up attempts to explain the outcomes (2), (3) had been presented in ref. [8] and later reproduced in ref. [9] by C. Corda, who claimed that these results can yet be understood under the framework of the general theory of relativity (GTR), if (as he posits) a so-called additional effect of the synchronization of a clock placed at the origin of a rotating system with a clock situated in the laboratory frame is taken into account. According to Corda, this "synchronization effect" gives an additional contribution to the measured energy shift between the source of resonant radiation and the detector of radiation amounting to the relative value

$$\left(\frac{\Delta E}{E}\right)_{synchr}=-\frac{1}{6}\frac{u^2}{c^2}, \tag{4}$$

which (according to him) should be added to the relative energy shift between the source and the absorber due to the usual time dilation effect; i.e.,

$$\left(\frac{\Delta E}{E}\right)_{dilation} = -\frac{1}{2}\frac{u^2}{c^2}. \tag{5}$$

Hence, the total (measured) energy shift, as per Corda's contrivance, is defined as the summation of eqs. (4), (5); which, via comparison with eq. (1), yields

$$k=2/3, \tag{6}$$

thus seemingly being in agreement with the experimental results (2), (3). Based on this artifice, by Corda's logic, our experiments [1-4] represent nothing else but, remarkably enough, a "*new proof of Einstein's general theory of relativity*" [8-10].

Be that as it may, the fact remains that we already reported in our papers [11-13] the obvious errors taking place in refs. [8, 9].

In the meantime, Corda published a new paper [14] where, at last, he comes to realize the presence of an "important mistake" (in his own words) with regards to his previous "clock synchronization" derivation in refs. [8, 9]. However, such an "important mistake", as admitted by Corda, lies beyond his severe computational errors which we had indicated in ref. [13], and which had been ignored by Corda yet again in [14]. In this respect, we choose to leave C. Corda to think more about his previous failures that had been pointed out in [13]. Herein, we disclose new errors committed by Corda in his latest attempt [14] to reinstate his "synchronization effect" (section 2). Having eliminated these errors, we arrive, once again, at the strict equality between the proper time of the detector and the proper time of the clock situated at the origin of the rotating system; which leaves no room for any illusory "synchronization effect" whatsoever.

The latter result, in fact, finalizes any further discussions on this subject. Moreover, as we had shown in refs. [11, 12], a hypothetical "synchronization effect", even if it ever existed, could not be measured in Mössbauer rotor experiments in principle; wherefore the newfangled attempt [14] by Corda aimed to invalidate this conclusion once more demonstrates his persistent misunderstanding of the Mössbauer effect methodology as we shall have to yet again explain under section 3. We thereafter conclude in section 4.

## 2. Mössbauer rotor experiments and "the clock synchronization effect" by C. Corda

In this section, for the convenience of the readers, we first reproduce the original derivation of the so-called "synchronization effect" by C. Corda [8, 9]. Then, we indicate his errors in said derivation and furnish the correct result from ref. [13], which shows the exact coincidence of the proper time of the detector with the coordinate time at any rotational frequency. However, this result, which indicates the absence of any "synchronization effect" between a resonant source at the origin of a rotating system and a detector of $\gamma$-quanta, was discounted by Corda in his final paper [14], where he pushed his calculations to once more derive his "synchronization effect" – though in another way than that in [8, 9]. Even so, we show below that this latest derivation of "synchronization effect", as exercised in [14], is again mistakable, so that it adds nothing to the explanation of the origin of the EES.

In order to indicate explicitly the errors by Corda in [8, 9], we first of all remind that, following Ashby [15], he adopted the transformation between an inertial frame and a rotating frame in cylindrical coordinates in the form of

$$t = t', \; r = r', \; \phi = \phi' + \omega t', \; z = z', \tag{7a-d}$$

with $\omega$ being the angular velocity. Hereinafter, the non-primed quantities are associated with the laboratory frame, while the primed quantities are associated with the rotating frame.

Next, using the expression for the space-time interval in cylindrical coordinates in an inertial (laboratory) frame

$$ds^2 = c^2 dt^2 - dr^2 - r^2 d\phi^2 - dz^2, \tag{8}$$

and combining eqs. (7) and (8), we arrive at the following expression for the space-time interval in the rotating frame (the Langevin metric) [15, 16]

$$ds^2 = \left(1 - \frac{r'^2\omega^2}{c^2}\right)c^2dt'^2 - 2\omega r'^2 d\phi' dt' - dr'^2 - r'^2 d\phi'^2 - dz'^2, \qquad (9)$$

along with the corresponding expression for the proper time increment [15]

$$d\tau^2 = dt'^2\left[1 - \left(\frac{\omega r'}{c}\right)^2 - \frac{2\omega r'^2 d\phi'}{c^2 dt'} - \left(\frac{d\sigma'}{cdt'}\right)^2\right]. \qquad (10)$$

Here, for brevity, we designated $d\sigma'^2 = dr'^2 + (r'd\phi')^2 + dz'^2$.

Applying eq. (10) to the analysis of Mössbauer experiments in a rotating system, Corda followed Ashby [15] and neglected the terms of the second order in the ratio ($\omega r'/c$). Hence, he obtained

$$d\tau \approx dt' - \frac{\omega r'^2 d\phi'}{c^2}. \qquad (11)$$

Further on, Corda pointed out that the detector of $\gamma$-quanta, being at rest in the laboratory frame, is moving on a circular orbit with respect to the origin of a rotating frame, whereupon the related motion of the detector is described by the equality

$$d\phi' = \omega dt'. \qquad (12)$$

Thus, substituting eq. (12) into eq. (11), he landed at

$$d\tau = dt'\left(1 - \frac{r'^2\omega^2}{c^2}\right), \qquad (13)$$

which, according to Corda [8, 9], "*... represents the proper time increment $d\tau$ on the moving clock having radial coordinate r' for values v<<c*."

However, as we have shown in [13], both equations (11) and (12) are erroneous.

Firstly, the linear approximation (11) to the ratio ($\omega r'/c$) – which was warranted for a special problem considered in [15] – is obviously inapplicable to Mössbauer rotor experiments, where the measurement of the coefficient $k$ in eq. (1) implies the accuracy of calculations of at least the order $(\omega r'/c)^2$. Therefore, in applying eq. (10) to the analysis of these experiment, we must keep the terms $(d\sigma'/cdt')^2$ and $(\omega r'/c)^2$; whereby their omission by Corda in eq. (11) is illegitimate. This already indicates that the application of eq. (11) to Mössbauer rotor experiments is invalid.

Secondly, eq. (12) is also false, because the actual motional equation of the detector, as seen by an observer at the origin of the rotating frame, should be written as

$$d\phi' = -\omega dt', \qquad (14)$$

which results from eq. (7c) at $\phi = const$ (because the detector is at rest in the laboratory frame).

Thus, the papers by Corda [8, 9] contain pivotal mistakes. It is all the more so, since Corda considered the detector to be positioned just outside the rim of the rotor adjacent to the rotating absorber, which immediately entails that – although mathematically impermissible from the get-go as long as the speed of light can be exceeded – effectively placing the detector at a far enough $r$ (say, at a location of merely around few hundred kilometers away), thereby making sure that one has an ultra-relativistic tangential velocity, Corda's "synchronization effect" will lead to an infinite (!) time dilation. This already by itself is sufficient to demonstrate the absurdity of Corda's "clock synchronization" idea.

In our paper [13], we have emphasized that, in the analysis of Mössbauer rotor experiments, the exact expression for the proper time increment (10) must be used instead of the approximate eq. (11). Besides, the erroneous equation (12) must be replaced by the correct equation (14). Thus, substituting eq. (14) into eq. (10), we straightforwardly obtain

$$d\tau = dt', \qquad (15)$$

which means that the laboratory clock and the clock at the rotational axis remain synchronized with each other at any angular rotational frequency $\omega$, and the alleged "synchronization effect" by Corda completely disappears. This is quite understandable, because the center of the disk bears zero tangential velocity at any $\omega$.

Despite all this, in his recent paper [14], Corda still continues to ignore the fact that the equality (15) clearly betokens the absence of any "synchronization effect" between the clock at the origin of the rotating system and the laboratory clock, and tries to remanufacture his derivation of the so-called "synchronization effect" whilst recognizing the presence of "an important mistake" in his previous calculations [8, 9].

Nevertheless, as before, Corda does not comprehend that eq. (11), resulting from the linear approximation to eq. (10) with respect to the ratio ($\omega r'/c$), is inadmissible in the analysis of Mössbauer rotor experiments, and again reproduces this equation in ref. [14]. At the same time, he tacitly modifies the physical meaning of eq. (11) – as introduced in his first paper [8] on this subject regarding the time rate of the clock attached to the detector of $\gamma$-radiation – and now entreats us to consider "*...light propagating in the radial direction, which implies* $d\phi'=dz'=0$."

Within these limitations, Corda went to derive a set of equations, correct only formally from the mathematical viewpoint (see eqs. (11)-(17) of [14]), where, however, he does not understand their physical meaning. Thus, our next goal is to provide the correct interpretation to the mentioned equations.

To this end, we first of all ought to emphasize that eqs. (11)-(17) of [14] are obtained under the constraint

$$d\phi'=dz'=0, \qquad (16a\text{-}b)$$

which corresponds to the propagation of light in the radial direction of the *rotating system*. This, however, is manifestly impossible in empty space, where the photons emitted from the origin of a rotating system propagate along a straight line in the laboratory frame, thus delineating the equalities $\phi$=*constant*, $z$=*constant*. Therefore, instead of eqs. (16a-b), we should in reality have

$$d\phi=0,\ dz=0. \qquad (17a\text{-}b)$$

Hence, combining eqs. (17a-b) and (7c-d), we get

$$d\phi'=-\omega dt',\ dz'=0 \qquad (18a\text{-}b)$$

in place of eqs. (16a-b) used by Corda in [14].

Thus, substituting eqs. (18a-b) into eq. (9), one arrives at

$$ds^2=c^2dt'^2-dr'^2.$$

Therefore, using the equality $ds$=0 for light propagation, one reaches the trivial result

$$dt'=\frac{dr'}{c}, \qquad (19)$$

which, along with the adopted transformations (7a-b), directly reflects the fact that, in empty space, photons propagate rectilinearly with the light velocity $c$ in vacuum $c$, exempt of any "synchronization effect" made up by Corda.

While it is not altogether impossible, from an overstretched technical viewpoint, to force photons to propagate in the radial direction of the rotating frame where the equalities (16a-b) used by Corda in ref. [14] might be fulfilled; nevertheless, for this purpose, despite the fact that Corda does not make any mention of it, one has to introduce into the scheme of Mössbauer rotor experiments, say, a thin guide for photons with length $r$ co-rotating with the rotor and physically joining the source to the absorber. In such a hypothetical case, for a laboratory observer, the photons inside the rotating guide would indeed propagate along a curved path and reach the radial coordinate $r$ later than those photons propagating along a straight line of the laboratory frame. Only thence might we expect the propagation time of photons inside the guide connecting the source with the absorber to be greater than the propagation time (19) of photons moving across a straight line in unhindered space for a laboratory observer. What is calculated by Corda, although he makes no allusion of it, is such a propagation time inside a rotating guide, and its difference from the propagation time (19), would then be furnished by his eq. (17) of [14].

Despite everything, one should realize that eq. (17) of [14] corresponds to a fictitious experimental configuration where, say, the source of resonant $\gamma$-quanta and the resonant absorber would be connected together via a prospective photon guide fastened onto the rotor – which, however, had so far never been realized in all of the known Mössbauer experiments in a rotating system (see, e.g., [1-4, 6, 17-21]). Moreover, the presence (or absence) of the photon guide between the source and absorber cannot affect the indication of a clock attached to the detector, and hence, eq. (15) remains in force, confirming the absence of any "synchronization effect" *à-la* Corda between the clock in the origin of a rotating system and a laboratory clock. Therefore, the present fabrication of Corda, like his previous fictions, has no significance in the explanation of the origin of the EES.

Even so, we still do not exclude the possible future appearance of further attempts by C. Corda to revive his otherwise untenable "synchronization effect", which he drove into the ground since 2015 (see, e.g., [8, 10, 9, 14]) in spite of our ongoing explanations of his many mistakes [11-13]. As a continuation of this tradition, in the present paper, we find and highlight new mistakes perpetrated by Corda in his novel derivation [14] of the alleged "synchronization effect". We additionally had shown in refs. [11, 12] that any kind of "synchronization effect" – even should it hypothetically exist – lies altogether beyond the measurement capabilities of every Mössbauer rotor experiment, so much so that in no way could such a contrivance explain the observed EES between emission and absorption lines.

Nevertheless, under section 3 of ref. [14] titled "Erroneous criticism of our approach", Corda again tries to perpetuate his position about the measurability of the "synchronization effect"; thus demonstrating once and for all his total misapprehension of the Mössbauer effect methodology. Below, we remind the basic methodological principles of Mössbauer spectroscopy, and then explicitly indicate the errors by C. Corda.

## 3. Corda's "synchronization effect" as just pure fiction: Insight from the methodological viewpoint

The impossibility to measure the "synchronization effect" of Corda in Mössbauer rotor experiments had already been explained in our papers [11, 12] via the straightforward analysis of various aspects of the Mössbauer effect methodology. Taking further into account the fact that, prior to the staging and execution of our experiments [1-4], the first co-author of the present paper had, during many years, remained one of the leading scientists in the Mössbauer effect methodology (see, e.g., refs. [22-41] on this subject), we were practically sure that our clear explication [11, 12] of the impossibility to measure the so-called "synchronization effect" – assuming that it existed in the first place – in Mössbauer spectroscopy would be sufficient to stop any further pointless discussions on this subject.

Yet it did not; Corda continued to publish more faulty papers like [10, 9, 14], and some others along the same line. Section 3 of the latest paper [14] is especially filled with confusing and absurd claims – all of which serve to demonstrate that, like before, Corda does not in the least understand the principles of Mössbauer spectroscopy.

There is no meaning to comment step by step on every single one of the erroneous claims by Corda in [14], and we will consider below only some important points, which highlight his mistakes. But, before embarking on this path, we shall, for the convenience of the readers, mention some basic principles of the Mössbauer effect methodology.

The Mössbauer effect is characterized by a very small ratio of the width of the resonant line $\Gamma$ to the energy of the resonant $\gamma$-quanta $E$; which, for the most popular isotope $^{57}$Fe used in Mössbauer spectroscopy, has a typical value of near $10^{-12}$, and this varies for different iron compounds within the same order of magnitude (see, e.g., [42]). In order to measure the shapes and positions of the resonant lines on the energy scale for iron-containing samples (which provides us with precious information about their local structure), these samples are irradiated by resonant $\gamma$-quanta coming from a $^{57}$Co source, which undergoes radioactive decay to the excited state of $^{57}$Fe. The resonant radiation emitted from this source then passes across the sample towards a suitable detector of $\gamma$-quanta aimed at measuring its intensity. In this configuration (which is named "transmission geometry"), the intensity of detected $\gamma$-radiation acquires the maximal val-ue when the resonant lines of the source and the absorber do not overlap with each other (e.g., no resonant absorption), whereas the intensity of detected $\gamma$-radiation achieves the minimal value when the positions of the resonant lines of the source and the absorber on the energy scale coincide with each other (e.g., maximal resonant absorption). In the real practice of Mössbauer spectroscopy, the source of resonant radiation is usually characterized by a single resonant line with a known shape and position on the energy scale, while resonant absorbers often contain several resonant lines. In order to measure these lines, one has to provide a modulation of the energy $E$ of the resonant $\gamma$-quanta of the source, and then to measure the intensity of the radiation passing across the absorber as a function of $E$. For this purpose, one can use the linear Doppler effect by providing a periodic oscillation of the source of resonant radiation along the line joining the source, absorber, and detector. In such a case, the variation of the energy $\Delta E$ of the $\gamma$-radiation is given by the known equation

$$\Delta E = \frac{v}{c} E \tag{20}$$

written in the linear approximation to the ratio ($v/c$), where $v$ is the velocity of the source. Eq. (20) indicates that at $\Delta E=\Gamma$ and $\Gamma/E \cong 10^{-12}$, we obtain $v \cong 0.3$ mm/s, and in common practice of Mössbauer spectroscopy, the energy shifts and the widths of the resonant lines indeed are usually expressed in velocity units via eq. (20).

The next important parameter of the resonant line is its height, which – after being normalized to the level of the background (measured at the time moments, when the resonant lines of the source and the absorber do not overlap) – is called the “resonant effect” and defined by the equation [42]

$$\varepsilon' = \frac{I_b - I_r}{I_b}. \tag{21}$$

Here $I_b$ is the intensity of the background, while $I_r$ is the intensity of the $\gamma$-radiation passing across the absorber under maximal resonant interaction.

The intensities $I_b$ and $I_r$ of the $\gamma$-quanta passed across the resonant absorber are measured with a suitable stationary detector placed behind the absorber. We should highlight the fact that such a detector is designed exclusively to fix, in laboratory time, the events of the registration of resonant $\gamma$-quanta without the exact evaluation of their energy; i.e., it operates as a counter of resonant $\gamma$-quanta. At the same time, one should take into account the fact that sources of resonant radiation often contain several $\gamma$-lines with different energies, which is particularly the case for iron-57 Mössbauer spectroscopy [42]. Under these conditions, the detector should be energy-sensitive to some extent in order to select the resonant line among other $\gamma$-lines emitted by the source. For this purpose, a relative energy resolution of about 10 % is quite sufficient for iron-57 Mössbauer spectroscopy – as was the case in our experiments [1-4] where a proportional detector, filled by Xenon, had been used.

Further, for iron-containing samples, a typical value of the resonant effect is about a few percent, which necessitates a long duration of measurement, in order to provide a high statistical quality of the Mössbauer spectra up until when the difference $I_b$-$I_r$ in eq. (21) exceeds by many times the statistical uncertainty in the determination of the level of the background $I_b$.

The value of the resonant effect (21) can be considerably increased for special samples enriched with the $^{57}$Fe isotope, whose concentration in the natural blend of iron isotopes is only 2.2 % [42]. For such enriched samples (where the relative content of $^{57}$Fe can be near 100 %), the value of the resonant effect (21) can attain a few tens percent. Under such circumstances, even the smallest relative energy shifts between emission and absorption resonant lines, lying in the range of the tiny values $10^{-14}\ldots10^{-12}$, are nevertheless capable of inducing considerable variations (from a few to some tens percent) in the intensity of the resonant $\gamma$-radiation, passing across the resonant absorber. This means that, using ordinary tools for the measurement of the intensity of resonant $\gamma$-radiation, we acquire the unique possibility of evaluating extremely small relative energy shifts between the resonant lines of the source and the absorber.

This is the most remarkable feature of the Mössbauer effect, which underlies its successful application to various branches of materials science, and which also has been used for the performance of Mössbauer experiments in a rotating system aimed to evaluate the second order Doppler effect and the associated relativistic dilation of time.

As is known (see, e.g. [43]), the linear Doppler shift between emission and absorption lines does not emerge in these experiments, and eq. (20) is replaced by eq. (1). Here, it is worth highlighting the fact that only resonant absorbers enriched by $^{57}$Fe had been used in Mössbauer rotor experiments, which allowed to achieve the variation of the intensity of the resonant radiation that passes across the absorber at the measurable level of a few percent even at a very small variation of the ratio $u^2/c^2$ in the range of $10^{-14}\ldots10^{-12}$; with the latter corresponding to the typical values for the tangential velocity $u$ achievable in modern rotor systems (e.g., up to 300 m/s).

These favorable aspects of the Mössbauer effect actually opened a realistic venue to measure the coefficient $k$ of eq. (1) with comparably high precision, which, according to the latest measurements [1-4], is framed by eqs. (2) and (3).

Be that as it may, Corda published his paper [8] where he claimed that the discrepancy between the standard relativistic prediction (5) – owing to the dilation of time in an orbiting absorber – and the measurements results (2), (3) could supposedly be explained by a thus far unaccounted-for "synchronization effect" between the clock at the origin of a rotating system and the clock of the detector of $\gamma$-quanta, purportedly yielding the additional component (4) for the relative energy shift between the source and the detector.

However, in our subsequent comment [11], we accentuated that the component of the energy shift (4) cannot be measured even in principle; for the ratio $u^2/c^2$, lying in the range $10^{-14}\ldots10^{-12}$, is at least eleven orders of magnitude (!) smaller than the typical energy resolution (about 10 %, as we have mentioned above) of an ordinary detector of $\gamma$-quanta utilized in Mössbauer rotor experiments. This means that the "synchronization effect" C. Corda concocted as a corollary of his eq. (4) represents nothing else but pure fiction and should have been rejected from the start.

In spite of the quite evident reality of our argumentation above – *which by itself totally invalidates the approach by Corda* – he, astoundingly enough, continued to wear down the topic, and published one more paper [10] where he already recognized the fact that the detector itself is insensitive to the energy shift component (4); while claiming that adding up eqs. (4) and (5) should remain correct for an observer located in a laboratory frame – wherein one has, to paraphrase Corda, "*the final output of the measuring*" [10]. He therefore tried to justify his position as follows: "*…a total energy shift measured by an observer located in the fixed detector of $\gamma$-quanta is different from the one measured by an observer located in the rotating resonant absorber…*". But, we have shown in ref. [12] that this assumption straightforwardly contradicts classical causality, because, according to Corda, two different observers should come out with different count numbers (!) of detected signals after the completion of a measurement run, which is manifest nonsense.

Nevertheless, in his recent paper [14], Corda argues against our conclusion about the incompatibility of his approach with classical causality; in particular, he claims that, in ref. [12], we have missed an "important" nuance between physical and coordinate time in a rotating system – which (again, in the opinion of Corda) prevents the violation of classical causality in the rotor experiment.

Specifically, Corda now asserts in [14] that both observers – one attached to the orbiting absorber (the frame K′), and the other attached to the resting detector (the frame K) – "*… measure the same number of pulses in different intervals of proper time. It is exactly this issue which generates the additional effect of clock synchronization*" [14].

In this respect, we emphasize that the fact of different intervals of proper time in the frames K′ and K is evident inasmuch as reflecting the time dilation effect for the orbiting absorber, which yields the following relationship for measured intensities of resonant radiation passing the resonant absorber:

$$I' = I\left(1+\frac{u^2}{2c^2}\right) \tag{22}$$

written to the accuracy $c^{-2}$, where $I'$ is the intensity measured in the frame K′, and $I$ is the intensity measured in K. If the "synchronization effect" by Corda is also taken into account, then eq. (22), according to him, should allegedly be modified to the form

$$I' = I\left(1+\frac{2u^2}{3c^2}\right) \tag{23}$$

Here, via eqs. (22), (23), we are trying to translate into a normal scientific language the subsequent equations (38)-(41) of ref. [14]; which are in and of themselves senseless to such an extent that it becomes difficult to provide any comment on them – though we will nonetheless make an effort to tackle this issue just below.

At any rate, there is no meaning to discuss eqs. (22) and (23) because neither of them have practical significance. Indeed, in both equations, the relative difference $(I'-I)/I$ has the order of magnitude $u^2/c^2$, which in the best case is near $10^{-12}$ or even smaller. At the same time, a typical relative statistical uncertainty in the measurement of the intensity $I$ in the Mössbauer rotor experiments [1-4, 6, 17-21] varies inside the range of 0.1…1 % – for example, in our latest experiment [4], it was about 0.5 %, which is more than ten orders of magnitude greater than the ratio $(I'-I)/I$ in eqs. (22) and (23). Therefore, it is indeed pointless to discuss which equation – (22) or (23) – is closer to reality; because, for both of them, the difference between $I'$ and $I$ lies far beyond any measurement capabilities. This also signifies that the "synchronization effect" professed by Corda – even should it hypothetically exist – is virtually immeasurable too. This is simply the result, which we already had emphasized a few years ago in ref. [11].

Next, we have the pleasure to provide our comments with respect to Corda′s eqs. (38)-(41) of ref. [14]. In order to clarify the meaning of the designations adopted in these equations, we first reproduce eqs. (36), (37) of [14]; i.e.,

$$\overline{N} = \int_0^{T_R} I(\tau)d\tau\,,\ \overline{N}' = \int_0^{T_L} I'(\tau)d\tau\,, \tag{24a-b}$$

where $\overline{N}$, $\overline{N}'$ are the total numbers of the detected $\gamma$-quanta for respectively the rotating and resting observers (which should be equal to each other according to the causality principle), $T_R$ is the proper time which is measured by the rotating observer, and $T_L$ is the proper time in the laboratory frame. By the way, with the involvement of Corda's "synchronization effect", the initial time moments of the measurements should come out different in both frames – which means that the lower integration limits cannot be the same in both equations (24a) and (24b). However, this is a tiny nuance in comparison with further senseless equations by Corda. In particular, he defines the total measurement times in both frames as

$$T_R \approx \tau_1\left(1-\frac{1}{2}\frac{v^2}{c^2}\right) \tag{25}$$

(eq. (38) of [14]), and

$$T_L \approx \tau_1\left(1-\frac{1}{2}\frac{v^2}{c^2}-\frac{1}{2}\frac{v^2}{c^2}\right) = \tau_1\left(1-\frac{2}{3}\frac{v^2}{c^2}\right) \tag{26}$$

(eq. (39) of [14]), where $\tau_1 \cong r/c$, with $r$ being the radial distance between the source and the detector. Hence, the difference

$$T_L - T_R \approx -\frac{\tau_1}{6}\frac{v^2}{c^2} \tag{27}$$

(eq. (40) of [14]), and the time dilation effect between the frames K and K′ is of the order

$$\left|T_L - T_R\right| \approx \left|\left(10^{-12}...10^{-13}\right)\right| \times \tau_1 \tag{28}$$

(see his unnumbered equation after eq. (41) of ref. [14]).

Reproducing these equations in the present paper, we get a strong impression that Corda actually does not understand what typical measurement time – which is denoted as $T_R$ in the frame K′, and $T_L$ in the frame K – is chosen in real measurements of the Mössbauer effect.

We especially highlight this point, because eqs. (38) and (39) of [14] (now eqs. (25) and (26), correspondingly) evince that, according to Corda, both $T_R$ and $T_L$ should have the order of magnitude comparable with $\tau_1 \cong r/c$; where the latter time interval in our experiment [4] was equal to 0.5 ns (at $r$=0.15 m). Does Corda really think that, during 0.5 ns, the intensity of resonant radiation passing across a resonant absorber can be measured? (For comparison, we ought to mention that the measurement time at each fixed rotational frequency in the experiment [4] was equal to 6000 s – i.e., more than 1.5 h).

Further, at $\tau_1$=0.5 ns, we obtain the numerical value of the difference

$$|T_L - T_R| \approx \left|\left(10^{-12}...10^{-13}\right)\right| \times \tau_1 \approx 0.5 \times \left|\left(10^{-21}...10^{-22}\right)\right| s. \qquad (29)$$

It is obvious for everybody that such a tiny difference (which is totally impractical for any laboratory scale experiment) cannot affect the measured intensities of resonant radiation in Mössbauer rotor experiments, and Corda does not in the least seem to understand this.

In fact, eqs. (38)-(40) of ref. [14] (now eqs. (25)-(27)) serve to solely demonstrate a complete misunderstanding of experimental physics by Corda.

What is more, continuing to discuss eqs. (38), (39) and (40) of ref. [14], Corda further writes: "*The fundamental issue is that, contrary to the claim of the authors of [18]* (now ref. [12]) *the time dilation effect between the frames K and K′ is NOT negligible".* Here Corda obviously fails to grasp how our claim about the negligible time dilation effect between the frames K and K′ refers only to any attempt towards a direct measurement of the difference between $T_R$ and $T_L$ in eq. (27), and the validity of this claim is well confirmed by the numerical estimation (29). At the same time, for the measurement of the Mössbauer effect in a rotating system, the time dilation effect between the frames K and K′ plays an important role insofar as engendering the energy shift between emission and absorption lines – hence leading to a measurable variation of the intensity of resonant $\gamma$-radiation passing through the absorber. Corda simply does not understand this principal difference in the manifestations of the time dilation effect, and continues to write: "*In fact, the authors of [18]* (now ref. [12]) *claim that their Mössbauer rotor apparatus detects a total time dilation effect*

$$\Delta\tau \approx -\frac{2\tau_1}{3}\frac{v^2}{c^2}, \qquad (42)$$

*which can be explained through their proper gravitational theory. But this total time dilation effect is of the same order of the time dilation effect between the frames K and K′ as it is shown by Eq. (40)* (now eq. (27)). *Thus, if the quantity of Eq. (40) is negligible, also the quantity of Eq. (42) must be negligible and this merely implies that the Mössbauer rotor apparatus should not work".*

In this respect, we have good news for Corda: Either "Mössbauer rotor apparatus", which we used in both of our experiments [1, 2] and [3, 4] worked well. The problem for Corda is that he continues to misunderstand that the time dilation effect for the orbiting resonant absorber induces the corresponding shift of its resonant line according to eq. (1), which leads to a quite measurable (at the level of a few percent) variation of the intensity of the resonant $\gamma$-radiation passing across the absorber thanks to the Mössbauer effect. At the same time, any attempt to directly measure the difference between the proper time for the resonant absorber $T_R$ and the proper time in the laboratory frame $T_L$ due to the same time dilation effect is quite hopeless, which is well indicated by eq. (29).

The present analysis decisively demonstrates the principal impossibility to measure the so-called "synchronization effect" by Corda, even it had existed, and we hope it actually ends permanently all discussions on the subject. In this respect, we see no point in commenting further on other erroneous claims by Corda under section 3 of [1], which simply result from his alarming dearth of understanding of the Mössbauer effect.

Another issue is that, at the very beginning and at the very end of section 3 of [14], Corda advocates some unproven and false claims against our papers [12, 13, 43, 44], so that we have to react. A detailed response to all of these claims falls outside the scope of the current paper and shall be done elsewhere. Here, we feel the need to shortly clarify two items.

First of all, it must be emphasized that the gravitational theory by the Yarman-Arik-Kholmetskii scientific collaboration (abbreviated as YARK theory for easy referencing) actually predicts the value $k$=2/3 in eq. (1) without introducing any "synchronization effect" *a-la* Corda, but based instead on a quantum mechanical description of resonant nuclei inside the absorber′s cells [43]. It is remarkable to notice that YARK theory is based on the original approach by T. Yarman suggested long before the experimental undertakings [1-4] (see, e.g., ref. [5]). At the same time, even in the face of this and other impressive successes of YARK theory towards the explanation of both old and modern cosmological observations (see, e.g., [44]), we did not, as Corda claims, wholesale "*…insinuate that such a new theory should replace GTR as the correct theory of gravity…"* Rather, we keep on proposing that YARK theory deserves attention from the scientific community due to its incontestable successes in explaining numerous empirical facts, and amongst other things, all of the end results of GTR; and that YARK theory, we hope, shall be considered useful in further advanced attempts to harmonize gravitational theory with quantum mechanics – given that YARK is already in symbiosis with quantum mechanics.

Next, finalizing section 3 of [14] Corda writes: "*We note that the authors of [17, 18]* (now refs. [43] and [12]) *recently published a new work with a further, clumsy attempt to show that our results on the Mössbauer rotor experiment are wrong [30]* (now ref. [13]). *The present paper shows that they are the results in [30] which are wrong instead".*

Given that Corda did not provide any explanation as to why, in his opinion, the results of our paper [13] "are wrong", we remind the readers that our paper [13] clarifies two principal points:

1. Equation (11), representing a linear approximation to the proper time interval in the Langevin metric (10) with respect to the ratio ($\omega r'/c$), is inapplicable to Mössbauer rotor experiments, dealing with the energy shifts between resonant lines in the higher order $(\omega r'/c)^2$.

2. Using the exact equation for the proper time (10), as well as the correct motional equation of the detector for an observer in a rotating frame (eq. (14)), we arrive at eq. (15), which shows the exact coincidence of the proper time of the detector with coordinate time; and this outcome leaves no room for insinuations about any "synchronization effect".

We continue to remain certain that both these statements of ref. [13] are fully correct.

## 4. Conclusion

We have shown that the new attempt by C. Corda towards the re-derivation of his "synchronization effect" between a clock at the origin of the rotating system and a laboratory clock implies a wholly unrealistic situation where the source at the origin of a rotating system and the absorber on the rotor rim should, for instance, be connected via a photon guide rigidly clamped onto the rotor; whereas Corda does not make mention of any of these.

In any case, such a configuration, where the photons are forced to propagate in the radial direction of a rotating frame, has never been realized in Mössbauer rotor experiments up to date. Therefore, the corresponding eq. (16a) employed by Corda for the description of such a motion of photons is totally unreal. Furthermore, the presence or absence of the photon guide between the source and absorber cannot affect the proper time of the detector, and hence, our eq. (15) remains in force anyway, confirming the absence of any "synchronization effect" *à-la* Corda between the clock in the origin of a rotating system and a laboratory clock. Therefore, the present fabrication by Corda just like his previous fictions, has no significance in the explanation of the origin of the EES as disclosed through the experiments [1-4], conducted by our team.

One must recall that, in real experiments (including [1-4, 6, 17-21]), $\gamma$-quanta emitted by a source freely propagated along the straight lines of the laboratory frame, where eq. (16a) used by Corda should be replaced by eq. (18a), along with the further derivation of eq. (19). The latter equation reflects the simple fact that, in empty space, photons propagate with the light velocity $c$ in vacuum exempt from any "synchronization effect" as imagined by Corda.

Therefore, any further discussion about any would-be contribution of the so-called "synchronization effect" to the measured energy shift between the resonant lines of a source at the origin and an absorber on the rotor rim is not even required.

Notwithstanding, in the current paper, we have once again shown (see also refs. [11, 12]) that Corda's so-called "synchronization effect" – even if it could be hypothetically assumed to exist at all – is totally immeasurable in Mössbauer rotor experiments. This is so much so that the obdurate contumacy displayed by Corda against this conclusion has no merit and only discloses his acute lack of understanding of the Mössbauer effect methodology.